\documentclass[twocolumn,superscriptaddress,amsfont,amssymb,amsmath, showpacs,balancelastpage, nofootinbib]{revtex4-1}
\usepackage{graphicx,longtable,mathrsfs,color,array}
\usepackage[unicode=true,pdfusetitle,
bookmarks=true,bookmarksnumbered=true,bookmarksopen=true,bookmarksopenlevel=1,
breaklinks=false,pdfborder={0 0 0},backref=false,colorlinks=true]{hyperref}
\hypersetup{citecolor=blue,filecolor=blue,linkcolor=blue,urlcolor=blue}
\usepackage[usenames,dvipsnames]{xcolor}
\usepackage{amssymb,amsmath,mathtools,mathrsfs}
\usepackage{epsfig,subfigure,placeins,float}
\usepackage{booktabs,longtable,ctable,multirow}
\usepackage{exscale,relsize}
\usepackage[normalem]{ulem}
\usepackage[T1]{fontenc}
\usepackage[utf8]{inputenc}
\usepackage{enumerate}
\usepackage{times, mathptmx}

\newcolumntype{C}[1]{>{\centering\let\newline\\\arraybackslash\hspace{0pt}}m{#1}}

\begin{document}
\title{Strong constraints on cosmological gravity from GW170817 and GRB 170817A.}
\author{T.~Baker}
\affiliation{University of Oxford, Denys Wilkinson Building,
  Keble Road, Oxford, OX1 3RH,  UK}
  \author{ E.~Bellini}
\affiliation{University of Oxford, Denys Wilkinson Building,
  Keble Road, Oxford, OX1 3RH,  UK}
   \author{P.~G.~Ferreira}
  \affiliation{University of Oxford, Denys Wilkinson Building,
  Keble Road, Oxford, OX1 3RH,  UK}
  \author{M.~Lagos}
  \affiliation{Kavli Institue for Cosmological Physics, The University of Chicago, Chicago, IL 60637, USA}
 \author{J.~Noller}
  \affiliation{Institute for Theoretical Studies, ETH Zurich, Clausiusstrasse 47, 8092 Zurich, Switzerland}
  \author{I.~Sawicki}
  \affiliation{CEICO, Fyzikální ústav Akademie věd ČR, Na Slovance 2, 182 21 Praha 8, Czechia} 
\begin{abstract}
	The detection of an electromagnetic counterpart (GRB 170817A) to the gravitational wave signal (GW170817) from the merger of two neutron stars  opens a completely new arena for testing theories of gravity. We show that this measurement allows us to place stringent constraints on general scalar-tensor and vector-tensor theories, while allowing us to place an independent bound on the graviton mass in bimetric theories of gravity. These constraints severely reduce the viable range of cosmological models that have been proposed as alternatives to general relativistic cosmology.
 \end{abstract}

  \date{\today}
  \maketitle
\noindent \textit{Introduction:} The advanced Laser Interferometer Gravitational Observatory (aLIGO) and the VIRGO interferometer, have recently announced the detection of gravitational waves (GW170817) from the merger of a neutron star (NS) binary located near NGC 4993 \cite{PhysRevLett.119.161101}. A gamma ray burst (GRB 170817A), occurring within 1.7 seconds, and in the vicinity of GW170817, was observed by the \textit{Fermi} Gamma-ray Burst Monitor, and the Anti-Coincidence Shield for the Spectrometer for the \textit{International Gamma-Ray Astrophysics Laboratory} \cite{2041-8205-848-2-L14,2041-8205-848-2-L15}. There is strong evidence that this event is an electromagnetic counterpart to the NS-NS merger \cite{2041-8205-848-2-L13,2041-8205-848-2-L12}. Comparing the travel time of light and gravitational waves (hereafter GW), we can place stringent constraints on cosmological gravity, and cosmology more generally \cite{Amendola:2012ky,Amendola:2014wma,Linder:2014fna,Raveri:2014eea,Saltas:2014dha,Lombriser:2015sxa,Jimenez:2015bwa,Bettoni:2016mij,Sawicki:2016klv}. 

We will assume that constraints on Lorentz violation in the electromagnetic sector are sufficiently strong  that the speed of light is $c=1$. In vacuum, Lorentz symmetry implies that {\it all} massless waves propagate at the speed of light. However, when a medium is present, Lorentz symmetry is spontaneously violated and propagation speeds can differ. Alternative theories of gravity, directly coupling extra degrees of freedom to curvature, provide such a medium when the new degree of freedom takes a configuration that defines a preferred direction (such as the time direction in cosmology). The action for linearized gravitational waves in such a medium takes the form:
\begin{eqnarray}
S_h=\frac{1}{2}\int d^3x dt \,M_*^2\left[{\dot h}_A^2-c^2_T(\nabla h_A)^2\right]\,. \label{eq:GWaction}
\end{eqnarray}
We have decomposed the metric as $g_{\alpha\beta}=\eta_{\alpha\beta}+h_{\alpha\beta}$ --- with $\eta_{\alpha\beta}$ the Minkowski metric --- by choosing locally inertial coordinates with time chosen to be the direction defined by the medium. We have expanded
	$h_{\alpha\beta}$ in polarization states, $\epsilon_A$, with amplitudes $h_A$, where $A=\times,+$. $M_*$ is the effective Planck mass, which in media provided by alternative gravity theories can differ from the standard $M_\text{P}$. $c_T$ is the speed of gravitational waves; we will find it convenient to parametrize this as \cite{Bellini:2014fua},
\begin{eqnarray}
c^2_T=1+\alpha_T \,.
\end{eqnarray}
In principle, $\alpha_T$ could adopt either positive or negative values. However, negative values ($c_T<c$) are constrained to $\alpha_T>-10^{-15}$ by a lack of observed gravi-\v{C}erenkov radiation from cosmic rays \cite{Moore:2001bv}. Up to now, the only upper bound on the \emph{propagation speed} of GWs comes from measuring the travel time between the two detectors of aLIGO, and is $\alpha_T<0.42$ \cite{Blas:2016qmn,Cornish:2017jml}.

In the regime we are considering (a gravitational wave propagating in effectively empty space, other than the medium provided by the new degree of freedom) the linearized action (\ref{eq:GWaction}) is sufficient. It is conceivable (but unlikely) that there may be some exotic behavior close the GW sources, in regions of strong gravity (for example, as occurs with the screening of scalar forces) that leads to non-linear corrections. Such effects could alter GW production, but will have no bearing on the gravitational wave propagation during the bulk of its travel time. Also, though Eq.~(\ref{eq:GWaction}) is valid for a wide range of gravitational theories, it does not encompass bimetric theories. \vspace{1mm}

\noindent \textit{Constraint on tensor speed excess:} 
We consider the geometric optics limit of Eq.~(\ref{eq:GWaction}) so that $c_T$ is indeed the speed of gravitational waves.

 Let $t_s$ be the time of emission for both the gravitational waves and photons; there can be some delay between the two emission peaks (for a NS-NS merger this will be a few seconds, of the order of the GRB duration), but this would only serve to tighten the bound we derive below. Let $t_T$ be the merger time identified in the gravitational wave train, and $t_c$ be the measured peak brightness time in the optical signal. To good accuracy, then, the transit time of the GW and photon signals are $c_T\left(t_T - t_s\right)=d_s$ and  $\left(t_c - t_s\right)=d_s$, 
where $d_s\simeq 40$ Mpc is the distance to the source. We then have 
${(t_c - t_T)}/{d_s} =1-{1}/{c_T}$.
Taylor expanding this gives $
\alpha_T\simeq {2\Delta t}/{d_s} 
$,
where $\Delta t$ is the measured difference in arrival times between the GW merger peak and the peak optical brightness. An arrival delay of $\Delta t \simeq 1.7$ seconds implies that
\begin{eqnarray}
\left|\alpha_T\right|\lesssim 1\times10^{-15}. \label{eq:constrGRB}
\end{eqnarray}
Comparing this to current cosmological constraints (where $\sigma_{\alpha_T}\sim 1$ \cite{HiCLASS}) or forecast cosmological constraints (where $\sigma_{\alpha_T}\sim 0.1$ \cite{Alonso:2016suf}), this constraint is remarkable. For all intents and purposes, we will hereafter consider $\alpha_T\simeq 0$ and attempt to understand its consequences for cosmology.\vspace{1mm}

\noindent \textit{Implications for scalar-tensor theories:} 
We begin by considering scalar-tensor theories. The Horndeski action is the most general scalar-tensor theory with second-order equations of motion \cite{Horndeski:1974wa,Deffayet:2011gz}, and is given by
$
S=\int \!\mathrm{d}^4x \sqrt{-g}\left\{\sum_{i=2}^5{\cal L}_i[\phi,g_{\mu\nu}]+{\cal L}_M[g_{\mu\nu},\cdots]\right\},
$ 
where ${\cal L}_M$ is the minimally coupled matter action. The scalar field lagrangian is built of four terms: two minimally coupled to gravity, ${\cal L}_2= K$ and ${\cal L}_3=  -G_3 \Box\phi $ and two terms explicitly involving the Ricci curvature, $R$, and the Einstein tensor, $G_{\mu\nu}$: 
\begin{eqnarray}
{\cal L}_4&=&   G_4R+G_{4,X}\left\{(\Box \phi)^2-\nabla_\mu\nabla_\nu\phi \nabla^\mu\nabla^\nu\phi\right\}  , \nonumber \\
{\cal L}_5&=& G_5G_{\mu\nu}\nabla^\mu\nabla^\nu\phi
-\frac{1}{6}G_{5,X}\big\{ (\nabla\phi)^3
-3\nabla^\mu\nabla^\nu\phi\nabla_\mu\nabla_\nu\phi\Box\phi 
 \nonumber \\ & & 
+2\nabla^\nu\nabla_\mu\phi \nabla^\alpha\nabla_\nu\phi\nabla^\mu\nabla_\alpha \phi
\big\}   \,.
\end{eqnarray}
Here $K$ and $G_i$ are functions only of $\phi$ and $X\equiv-\nabla^\nu\phi\nabla_\nu\phi/2$, and subscript commas denote derivatives. 
On a cosmological background, Horndeski models give \cite{DeFelice:2011bh}
\begin{eqnarray}
M^2_*\alpha_T&\equiv&2X\left[2G_{4,X}-2G_{5,\phi}-\left(\ddot{\phi}-\dot{\phi}H\right)G_{5,X}\right] \label{eq:alphaTH}
\end{eqnarray}
where $
M^2_*\equiv2\left(G_4-2XG_{4,X}+XG_{5,\phi}-{\dot \phi}HXG_{5,X}\right)
$.

One way of satisfying $\alpha_T\sim0$ is through a delicate cancellation between $G_{4,X}$, $G_{5,\phi}$ and $G_{5,X}$. If $G_{5,X}=0$, this cancellation is trivial, since it implies that the theory is minimally coupled. Any non-trivial cancellation would not only have to be time-dependent, but also sensitive to the matter content of the universe due to the dependence on $H$ and $\ddot{\phi}$. Thus even a small change in e.g.\ the dark matter density, or deviations from isotropy and homogeneity, would severely violate it.  
Furthermore, any such a cancellation would be accidental, with no symmetry to protect it. Some shift symmetric Horndeski actions (i.e. not dependent on $\phi$) are, to some degree, stable to radiative corrections. In \emph{flat} spacetime, for $K,G_i$ linear in $X$ (galileons \cite{Nicolis:2008in}), there exists an exact quantum non-renormalization theorem \cite{Luty:2003vm,Goon:2016ihr,Saltas:2016awg} --- there are no corrections to these operators. The corrections remain under control when the galilean symmetry is weakly broken \cite{Pirtskhalava:2015nla}, as it must be in curved spacetime. In this case, the Horndeski interactions are suppressed by a scale $\Lambda_3$, whereas quantum corrections enter suppressed by the parametrically larger scale $\Lambda_2 \gg \Lambda_3$, which satisfies $\Lambda_2^4 = M_{\rm Pl} \Lambda_3^3$ \cite{Pirtskhalava:2015nla}. 
A typical value is $\Lambda_3 \sim 10^{-13} $eV, leading to $\Lambda_3/\Lambda_2 \sim 10^{-10}$. 
With relatively mild assumptions on the $G_i$ functions, this can be shown to lead to order $(\Lambda_3/\Lambda_2)^4 \sim 10^{-40}$ corrections on the $G_i$ \cite{Creminelli:2017sry} (and hence on $\alpha_T$ as derived here), showing that a $10^{-15}$ tuning of $\alpha_T$ unspoiled by radiative corrections can be achieved.

A more natural interpretation of the constraint \ref{eq:constrGRB} is that \emph{each} of the terms ($G_{4,X}$, $G_{5,\phi}$, $G_{5,X}$) contributing to $\alpha_T$ is zero, i.e.\ that ${\cal L}_5\propto G_{\mu\nu}\nabla^\mu\nabla^\nu\phi$, vanishing identically as a result of the Bianchi identity, while ${\cal L}_4=f(\phi)R$, i.e.\ the coupling to gravity can at most be of the Jordan-Brans-Dicke (JBD) type. Such a restriction reduces the viable model space for scalar-tensor modified gravity to two classes: (i) models in which the scalar does not evolve significantly on cosmological timescales, and (ii) those in which it does. 
	
Class (i) is the generalized JBD class, including models such as $f(R)$ gravity. Such models require chameleonic screening to evade solar-system tests of gravity, and therefore cannot have a background evolution significantly different from that of concordance cosmology; they do not self-accelerate cosmological expansion \cite{Wang:2012kj}. The sound speed of the scalar fluctuations is equal to that of light. On the other hand, the strength of the fifth force, $f_{,\phi}$, is allowed to be similar to gravity. 
	
In Class (ii), the scalar evolves quickly, $X\sim H^2M_*^2$, and non-canonical kinetic terms in $G_2$ and $G_3$ play a significant role: they can give rise to self-acceleration, significantly changing the equation of state and the sound speed. Constraints on the evolution of the Planck mass  \cite{Uzan:2010pm,Ade:2014zfo} restrict the strength of coupling to gravity $f_{,\phi}$ to be small, since the scalar runs during the entire history of the universe in these models. We reiterate that perturbative control of quantum corrections in the fast-moving models depends on shift symmetry, which would disallow any dependence on $\phi$ in the action, specifically the conformal coupling $f(\phi)$. 

Horndeski theory is not the most general scalar-tensor theory propagating one single extra degree of freedom. New terms can be added to construct the ``beyond'' Horndeski Lagrangian \cite{Gleyzes:2014dya,Zumalacarregui:2013pma} at the price of third derivatives in equations of motion and new constraints to remove any extra degrees of freedom naively implied by them. This extension is described by two new free functions, $\tilde{G}_4\left(\phi,\,X\right)$ and $\tilde{G}_5\left(\phi,\,X\right)$ correcting ${\cal L}_{4}$ and ${\cal L}_{5}$ (see \cite{Gleyzes:2014dya} for the complete expressions) and modifying Eq.~(\ref{eq:alphaTH}) to
\begin{align}\label{eq:bH_alphaT}
\alpha_{\textrm{T}}M_{*}^{2}&=  4X\left(G_{4,X}-\tilde{G}_{4,X}-G_{5,\phi}\right)-2\ddot{\phi}XG_{5,X}\nonumber\\
&\qquad+2\dot{\phi}HX\left(G_{5,X}-\tilde{G}_{5,X}\right)\,,
\end{align}
where $M_{*}^{2}= 2G_{4}-4X\left(G_{4,X}-\tilde{G}_{4,X}\right)+2XG_{5,\phi}-2\dot{\phi}HX\left(G_{5,X}-\tilde{G}_{5,X}\right)$.

It is clear from Eq.~(\ref{eq:bH_alphaT}) that one option is to set all the terms contributing to $\alpha_T$ to zero, as in the Horndeski case. An intriguing alternative is to choose $G_{5,X}=\tilde{G}_{5,X}=0$ and
$\tilde{G}_{4,X}=G_{4,X} -G_{5,\phi}$, which indeed leads to 
 $\alpha_T=0$ but also allows for $M_*\neq M_P$ and $\alpha_H\neq 0$, where $\alpha_H$ is the additional beyond-Horndeski parameter introduced in \cite{Gleyzes:2014dya}. Although it is beyond the scope of this work to discuss the properties of this particular model, we should emphasize that this is the only algebraic choice for the $G_i$ functions that ensures $\alpha_T=0$ regardless of the underlying cosmology.

In our discussion of scalar-tensor theories, we should briefly mention Degenerate Higher-Order Scalar-Tensor (DHOST) theories \cite{Langlois:2015cwa,BenAchour:2016fzp}. DHOST theories are constructed to be a further generalization of Horndeski, but have to include new constraints  to  avoid Ostrogradsky instabilities. The result is a long list of classes of theories ($\simeq30$) having disjoint parameter spaces, but which on a cosmological background reduce to just two types \cite{Langlois:2017mxy}. One is unstable and thus irrelevant here. The other can be transformed to beyond Horndeski with a conformal transformation of the form $\tilde{g}_{\mu\nu}=C(X)g_{\mu\nu}$. Conformal transformations leave null geodesics null. Thus if a DHOST model describes gravity in cosmology, then the requirements for $\alpha_T=0$ listed above apply to the beyond-Horndeski counterpart of the DHOST theory.

To conclude, if we assume that it is not possible to enforce precise cancellations for the reasons discussed above, the constraint on $\alpha_T$  excludes such models as the quartic and quintic galileon or a generic beyond Horndeski, leaving only models which are conformally coupled to gravity. On the other hand, models where gravity remains minimally coupled remain unconstrained: fast-moving models such as kinetic gravity braiding \cite{Deffayet:2010qz} can give rise to self-acceleration and admit an interpretation as the dynamics of a superfluid \cite{Pujolas:2011he}, rather than as a modification of gravity. Finally, quintessence models remain unconstrained. 

\noindent \textit{Implications for  vector-tensor theories:} 
We now turn to vector tensor theories of gravity, i.e.~theories where the additional gravitational degree of freedom is given by a 4-vector, $A^\mu$. First, we consider Generalized Einstein-Aether gravity, where $A^\mu$ is time-like and the action is 
$
S=\int d^4x\sqrt{-g}\left[\frac{M^2_P}{2}R+{\cal F}(K)+\lambda(A^\mu A_\mu+1) \right]$,
where $\lambda$ is a Lagrange multiplier, $K=c_1\nabla_\mu A_\nu \nabla^\mu A^\nu+c_2(\nabla_\mu A^\mu)^2+c_3\nabla_\mu A_\nu \nabla^\nu A^\mu$ (with $c_i$ constants) and ${\cal F}(x)$ is an arbitrary function \cite{Jacobson:2000xp,Zlosnik:2006zu}. In this model $\alpha_T=-(c_1+c_3){\cal F}_{,K}/[1+(c_1+c_3){\cal F}_{,K}]$, so the constraint on $\alpha_T$ implies $c_1=-c_3$. 
On Minkowski space, this reduces the theory to the Maxwell action (with a time-like constraint). On a cosmological background, we still allow for modifications as $3M^2_PH^2=(\rho-{\cal F}/2)(1-3c_2{\cal F}_{,K})$, whereas the effective Planck mass in Eq.~(\ref{eq:GWaction}), which is generally given by  $M^2_*=M^2_P[1-(c_1+c_3){\cal F}_{,K}]$, will reduce to the GR value. 

A second-class of vector-tensor theories of interest are generalized Proca theories \cite{Tasinato:2014eka,Heisenberg:2014rta}, whose 4D action is, much like Horndeski theory,  given by
$
S=\int d^4x \sqrt{-g} \left( {\cal L}
+{\cal L}_M \right),\,
{\cal L}={\cal L}_F+\sum_{i=2}^{5} {\cal L}_i\,,
$
where the vector field Lagrangian is built so that precisely one extra (longitudinal) scalar mode propagates in addition to the two usual Maxwell-like transverse polarisations. The individual $ {\cal L}_i$ are given by three minimally coupled terms, $
{\cal L}_F = -\frac14 F_{\mu \nu}F^{\mu \nu}$, ${\cal L}_2 = G_2(X)$ and ${\cal L}_3 = G_3(X) \nabla_{\mu}A^{\mu}$,  and two nontrivial terms given by
\begin{widetext}
\begin{eqnarray}
{\cal L}_4 &=&
G_4(X)R+
G_{4,X}(X) \left[ (\nabla_{\mu} A^{\mu})^2
+c_2 \nabla_{\rho}A_{\sigma} \nabla^{\rho}A^{\sigma}
-(1+c_2) \nabla_{\rho}A_{\sigma}
\nabla^{\sigma}A^{\rho} \right]\,,\nonumber \\
{\cal L}_5 &=&
G_{5}(X) G_{\mu \nu} \nabla^{\mu} A^{\nu}
-\frac16 G_{5,X}(X) [ (\nabla_{\mu} A^{\mu})^3
-3d_2 \nabla_{\mu} A^{\mu}
\nabla_{\rho}A_{\sigma} \nabla^{\rho}A^{\sigma}
-3(1-d_2) \nabla_{\mu} A^{\mu}
\nabla_{\rho}A_{\sigma} \nabla^{\sigma}A^{\rho}
\nonumber \\
& &
+(2-3d_2) \nabla_{\rho}A_{\sigma} \nabla^{\gamma}
A^{\rho} \nabla^{\sigma}A_{\gamma}
+3d_2 \nabla_{\rho}A_{\sigma} \nabla^{\gamma}
A^{\rho} \nabla_{\gamma}A^{\sigma}]\,.
\label{ProcaL}
\end{eqnarray}
\end{widetext}
As usual, $F_{\mu \nu}=
\nabla_{\mu}A_{\nu}-\nabla_{\nu}A_{\mu}$, $c_2$ and $d_2$ are constants,
and $G_{2,3,4,5}$ are arbitrary functions of $X=-\frac12 A_{\mu} A^{\mu}$. On a cosmological background $A^\mu=(A,{\vec 0})$ and $\alpha_T$ is given by 
\begin{eqnarray}
\alpha_T=A^2\left[2 G_{4,X} - (H A - {\dot A}) G_{5,X}\right]/q_T, \label{eq:alphaTProca}
\end{eqnarray}
where $q_T = 2 G_4 - 2A^2 G_{4,X} + H A^3 G_{5,X}$. 
Analogously to the scalar-tensor case considered above, if $\alpha_T = 0$ we either then have to carefully tune the functional dependence of $G_4$ and $G_5$ to satisfy this criterion (all the considerations about radiative stability, time dependence and background symmetry we discussed for Horndeski theories hold), or consider a theory with minimal higher-order interactions by requiring $G_{4,X} = G_{5,X} = 0$ leading to ${\cal L}_4 \propto R$ and ${\cal L}_5 \propto G_{\mu\nu}\nabla^\mu A^\nu$. In the latter case, ghost-freedom for tensor perturbations then enforces $G_4 > 0$, while ghost and gradient instabilities for vector modes are automatically satisfied.

In Generalised Proca theories the equation of motion for $A^\mu$ separates the evolution into two branches, one with a non-dynamical scalar degree of freedom and a second one with full dynamics for all three degrees of freedom, which we will focus on here. Requiring $G_{4,X} = G_{5,X} = 0$ (and hence $\alpha_T=0$) as above, the modified Friedman equation then becomes $
   3H^2 = (\rho-G_2)/(2 G_4)$, and thus
$2G_4$ describes a re-scaled constant Planck mass. We note that on the de Sitter fixed point of this model \cite{DeFelice:2016yws}, in the limit $\rho  = 0$, consistency will enforce $G_2 < 0$, due to the  ghost-freedom condition for tensor perturbations $G_4 > 0$.  

One can go a step beyond Generalised Proca theories and consider the "beyond" Generalised Proca  model of \cite{Heisenberg:2016eld} (also see \cite{Allys:2015sht,Jimenez:2016isa}). Here six new free functions enter at the level of the action, denoted $G_6, g_5, f_4,f_5,\tilde f_5,\tilde f_6$. Of the new functions only $f_4$ and $f_5$ affect the background evolution and that of linear tensor perturbations, whereas the remaining functions only affect linear vector and scalar perturbations. The $\alpha_T=0$ constraint now implies
$
G_{5,X} (HA-{\dot A}) - 2 G_{4,X} = 2 f_4 A^2 + 6 f_5 H A^3,
$ 
which depends on the new functions $f_4, f_5$. If we choose to set all participating functions to zero to ensure $\alpha_T = 0$, this means both the background and tensor perturbations will behave exactly as in the Generalised Proca case considered above. 

\noindent {\textit{Implications for bigravity  theories:} 
We now consider models with two coupled metrics. The only non-linear Lorentz invariant ghost-free possible interactions are given by the deRham-Gabadadze-Tolley (dRGT) potential  \cite{deRham:2010kj, deRham:2010ik, Hassan:2011zd}. The action is given by
$
S=\; ({M_g^2}/{2})\int d^4x\; \sqrt{-g}R_g +({M_f^2}/{2})\int d^4x\; \sqrt{-f}R_f 
- m^2M_{g}^2\int d^4x\; \sqrt{-g}\sum_{n=0}^4\beta_n e_n\left(\sqrt{g^{-1}f}\right),
$
where we have two dynamical metrics $g_{\mu\nu}$ and $f_{\mu\nu}$ with their associated Ricci scalars $R_g$ and $R_f$, and constant mass scales $M_g$ and $M_f$, respectively. Here, $\beta_n$ are free dimensionless coefficients, while $m$ is an arbitrary constant mass scale. The dRGT potential is defined in terms of the functions $e_n (\mathbb{X})$, which correspond to the elementary symmetric polynomials of the matrix $\mathbb{X}=\sqrt{g^{-1}f}$.

For simplicity, let us assume that matter fields are coupled minimally to the metric $g_{\mu\nu}$, and all the parameter $\beta$s are of order 1 . 

The bigravity action generally propagates one massive and one massless graviton; and the field $g_{\mu\nu}$ will be a combination of \textit{both} modes. The massless mode has a dispersion relation given by $E^2_0=k^2$, while the massive mode has $E_m^2=k^2+m^2$ (with omitted factors of $\beta$s of order 1) on Minkowski space (and a de-Sitter phase, i.e.~late times).  

Let us first discuss the restricted case of massive gravity, when $M_f/M_g\rightarrow \infty$, and only the massive graviton propagates (while the metric $f_{\mu\nu}$ is frozen). In this case, the dispersion relation of gravitational waves is $E^2=k^2+m^2$. As a result, the speed of GW will be frequency-dependent leading to a phase difference in the waveforms. Bounds from GW150914 led to $m \leq 1.2 \times10^{-22}$eV \cite{TheLIGOScientific:2016src}. 
With an EM counterpart to the GWs, the bound of 1.7 seconds on the time delay also leads to $m \lesssim  10^{-22}$eV (note that we have considered a frequency region of interest of $10-100$Hz and ignored the frequency dependency of the velocity, which is small) which is uncompetitive with Solar System
 fifth force constraints  of order $m \lesssim 10^{-33}$eV \cite{deRham:2016nuf}. 
In case of massive bigravity, assuming similar amplitudes for both modes, one has a fast oscillation with a slowly modulated amplitude. The frequency of the modulated wave is proportional to $m$ and hence negligible compared to the time scale of the NS merger. The dispersion relation of the fast mode is effectively that of a massive graviton $E^2=k^2+m^2$ (omitting again factors or order 1), and thus one obtains the same constraint as for massive gravity.

Unlike for scalar-tensor and vector-tensor theories, in massive gravity local constraints from GW propagation have no bearing on cosmology. In particular, the existence of scalar and tensor instabilities  \cite{Lagos:2014lca,Cusin:2014psa} in particular branches of the background cosmology will be unconstrained by the measurements discussed in this paper.
Further discussion on GW constraints in massive bigravity can be found in \cite{DeFelice:2013nba, Narikawa:2014fua,Max:2017flc}. Constraints in the case where both metrics are coupled to matter are discussed in \cite{Brax:2017hxh}.\vspace{0.5mm}

\noindent\textit{Caveats:}
We now address possible caveats. For a start, the source lies at a very low redshift ($z_s=0.01$); thus our constraint is on the speed of GWs \textit{today}. It would of course be a great coincidence if $\alpha_T$ were to vanish now with such precision, but not at other times. However, this is in principle a possibility.

Another uncertainty is the extent to which the effective metric relevant for the propagation of perturbations with wavelengths similar to the size of the universe, as studied in cosmology, is the same one experienced by the GW with the wavelength of 3000~km (to which aLIGO/VIRGO are sensitive). For cosmological modes with wavelengths of 10--100~Mpc, taking the background --- the medium in which fluctuations propagate --- to be isotropic and homogeneous is a good approximation. Wavelengths probed by aLIGO/VIRGO are much shorter than the typical size of structures in the universe, so the GW should be sensitive to the inhomogeneities. Indeed, one can argue that, apart from the initial exit from the source galaxy and the final entrance into the Milky Way, the GW was mostly propagating through space with density of matter significantly below the current cosmic average, when averaged over scales of the order of the GW's wavelength.

Some alternative theories of gravity depend crucially on a highly non-linear response to the matter density by the extra degrees of freedom (the need for screening on Solar-System scales). This may well mean that the GW speed predicted for an averaged cosmology, and that for the matter density along the particular trajectory this GW took could be different. Thus there would not be a simple connection between the time delay observed and the properties of gravity on cosmological scales. We would argue that, if such an effect is relevant, then the GW would be propagating with a speed which the cosmological modes will experience when the universe has emptied out to the same extent as the averaged density along the trajectory of the GW. If $\alpha_T$ is evolving, we may well have measured its asymptotic future value. 

\noindent\textit{Conclusions:} The detection of GW170817, together with its EM counterpart (GRB 170817A), bounds the speed of gravitational waves to deviate from $c$ by no more that one part in $10^{15}$. This single fact has profound repercussions for extended gravity models which are of interest in current cosmology. We summarize here the key consequences explained in this letter:

\noindent \textit{i)} Assuming no finely-tuned cancellations between Lagrangian functions occur, the only viable scalar-tensor theories have a gravitational action of the form $\propto f(\phi)R$ (plus non-gravitational terms), i.e.\ conformally coupled theories. This eliminates, for example, the quartic and quintic Galileons. Quintessence is still allowed as the minimally coupled limit of these theories.

\noindent \textit{ii)} In the conformally coupled class, the only surviving self-accelerating theories must have a small (or, indeed, minimal) coupling strength to gravity, and hence can be treated as generalized fluids. Models in this category include cubic Galileons, kinetic gravity braiding and k-essence.

\noindent \textit{iii)} The ``beyond'' Horndeski extension of scalar-tensor theories introduces only one further surviving model, which is also conformally coupled to gravity.

\noindent \textit{iv)} For vector fields, assuming no finely-tuned cancellations, (Generalized) Einstein-Aether models are now subject to the stringent relation $c_1 = -c_3$. 

\noindent \textit{v)}
``Beyond'' and standard Generalized Proca 
models,  assuming no finely-tuned cancellations, behave identically at background level, with vastly simplified higher order gravitational interactions, such as a coupling to $R$, where the proportionality constant acts as a rescaled Planck mass in the Friedmann equations.

\noindent \textit{vi)} In bimetric theories the mass of the graviton is constrained to be $m \lesssim 10^{-22}$ eV, which is weaker than current Solar System bounds but entirely independent of them. This constraint has no bearing on cosmology.

For the first time, powerful and  general statements can be made about the structure of (non-)viable gravitational actions, and some current popular models are ruled out (also see~\cite{Creminelli:2017sry,Sakstein:2017xjx,Ezquiaga:2017ekz,Wang:2017rpx}). These decisive statements will undoubtedly shape the direction of future research into extensions of General Relativity.
\vspace{-0.3in}
\section*{Acknowledgments}
\vspace{-0.2in}
\noindent We acknowledge conversations with Rob Fender, Filippo Vernizzi and Miguel Zumalacárregui, and the discussions made possible by the DARKMOD workshop at IPhT Saclay.  TB is supported by All Souls College, University of Oxford. EB is supported by the ERC and BIPAC. PGF acknowledges support from STFC, BIPAC, the Higgs Centre at the University of Edinburgh and ERC.  ML is supported at the University of Chicago by the Kavli Institute for Cosmological Physics through an endowment from the Kavli Foundation and its founder Fred Kavli.
JN acknowledges support from Dr.\ Max Rössler, the Walter Haefner Foundation and the ETH Zurich Foundation. IS is supported by ESIF and MEYS (Project CoGraDS -- CZ.02.1.01/0.0/0.0/15\_003/0000437). 
\bibliography{RefModifiedGravity}

\end{document}